\documentclass{article}

\usepackage{mathtools}
\usepackage{siunitx}
\usepackage{authblk}
\usepackage{xspace}

\newcommand{\RPBlong}{radially polarized beam\xspace}
\newcommand{\TFlong}{tightly focused\xspace}

\newcommand{\ETO}{E\textsubscript{\tiny{1}}(TO)\xspace}
\newcommand{\ATO}{A\textsubscript{\tiny{1}}(TO)\xspace}
\newcommand{\Eh}{E\rlap{\textsubscript{\tiny{2}}}\textsuperscript{\tiny{h}}\xspace}
\newcommand{\invcm}{\ensuremath{\,\mathrm{cm}^{-1}}}

\title{Towards Polarization-based Excitation Tailoring for Extended Raman Spectroscopy}

\author[1,2,3,$\dag$]{Simon Grosche} 
\author[1,2,$\dag$]{Richard H{\"u}nermann} 
\author[1,4]{George Sarau}
\author[1,4,5]{Silke Christiansen}
\author[6,7]{Robert W. Boyd}
\author[1,2,6]{Gerd Leuchs}
\author[1,2,6,*]{Peter Banzer}

\affil[1]{\small Max Planck Institute for the Science of Light, Staudtstr. 2, D-91058 Erlangen, Germany}
\affil[2]{Institute of Optics, Information and Photonics, Friedrich-Alexander-University Erlangen-Nuremberg, Staudtstr. 7/B2, D-91058 Erlangen, Germany}
\affil[3]{Current affiliation: Chair of Multimedia Communications and Signal Processing, Friedrich-Alexander-University Erlangen-Nuremberg, Cauerstr. 7, D-91058 Erlangen, Germany}
\affil[4]{Helmholtz-Zentrum Berlin f\"ur Materialien und Energie, Hahn-Meitner-Platz 1, D-14109 Berlin, Germany}
\affil[5]{Physics Department, Freie Universit\"at Berlin, Arnimallee 14, D-14195 Berlin, Germany}
\affil[6]{Department of Physics, University of Ottawa, 25 Templeton, Ottawa, Ontario, Canada K1N 6N5}
\affil[7]{Institute of Optics, University of Rochester, Rochester, NY 14627, USA}
\affil[$\dag$]{These authors contributed equally.}
\affil[*]{Corresponding author: peter.banzer@mpl.mpg.de}

\date{}   

\begin{document}

\maketitle

\begin{abstract}

Undoubtedly, Raman spectroscopy is one of the most elaborated spectroscopy tools in materials science, chemistry, medicine and optics. However, when it comes to the analysis of nanostructured specimens, accessing the Raman spectra resulting from an exciting electric field component oriented perpendicularly to the substrate plane is a difficult task and conventionally can only be achieved by mechanically tilting the sample, or by sophisticated sample preparation. Here, we propose a novel experimental method based on the utilization of polarization tailored light for Raman spectroscopy of individual nanostructures. As a proof of principle, we create three-dimensional electromagnetic field distributions at the nanoscale using tightly focused cylindrical vector beams impinging normally onto the specimen, hence keeping the conventional beam-path of commercial Raman systems. Using this excitation scheme, we experimentally show that the recorded Raman spectra of individual gallium-nitride nanostructures of sub-wavelength diameter used as a test platform depend sensitively on their location relative to the focal vector field. The observed Raman spectra can be attributed to the interaction with transverse or longitudinal electric field components. This novel technique may pave the way towards a characterization of Raman active nanosystems using full information of all Raman modes.
\end{abstract}

\section{Introduction}
Raman spectroscopy has become a widely spread and well established method. Key factors that enabled a huge variety of commercial and scientific applications of this technique are the availability of lasers providing narrow spectral line-widths at high powers in combination with highly sensitive spectrometers and interference filters capable of blocking the excitation wavelength while transmitting and measuring the Raman scattered light \cite{Smith2005}. The underlying physical effect of Raman spectroscopy is Raman scattering -- a highly inefficient inelastic scattering process --  with the incoming light interacting with the rotational and vibrational states of molecules, solids, liquids or gaseous media. Since the resulting Raman spectra are unique for different materials and nanostructured specimens, Raman spectroscopy has become a standard tool in material sciences, medicine and chemistry. Among many other applications, it can be used for identifying an unknown specimen, for example, in drug testing \cite{Penido2016}, for real-time and in-vivo disease diagnosis \cite{Hanlon2000}, for the measurement of temperatures \cite{Moore2014}, probing material properties such as strain \cite{Neumann2015,Sarau2014,Fluegel2015,Kang2005}, for the investigation of phase transitions \cite{Pinquier2004}, or even for determining the compounds (including minerals and water) on Mars \cite{Wang2003}.\\
However, an inherent disadvantage of Raman scattering is the typically ultra-small scattering cross section and conversion efficiency. Therefore, many different techniques have been developed in order to enhance the scattering signal. In this context, resonance Raman scattering \cite{Koningstein1988}, surface enhanced Raman scattering (SERS) \cite{LeRu2007}, hyper-Raman scattering \cite{Terhune1965} and tip enhanced Raman scattering (TERS) \cite{Sonntag2014} need to be mentioned. More advanced techniques have been proposed as well, including, among others, coherent anti-Stokes Raman spectroscopy (CARS) \cite{Maker1965,Begley1974,Tolles1977} and transmission Raman spectroscopy \cite{Waters1994}. Beyond methods aiming for an increase of the signal strength of Raman scattering, also other enhancements and improvements of Raman spectroscopy have been proposed. For instance, some implementations take advantage of the fact that Raman active modes are inherently polarization sensitive (see, e.g., Ref. \cite{Grzybowski1957,Damen1966}). By changing the input polarization direction of a linearly polarized excitation beam impinging normally on a specimen and by analyzing the polarization of the Raman scattering with another polarizer in forward, backward or lateral direction, more detailed information about the structure of a specimen, the orientation of molecules or the strain in materials under study can be retrieved. Nonetheless, for some measurements the polarization is required to be perpendicular to the surface of the sample or substrate to also get access to corresponding vibrational modes. With an incoming paraxial or weakly focused beam of light, this \textit{longitudinal} polarization component cannot be achieved for normal incidence since the longitudinal components of the electric field, i.e., the components oscillating along the propagation direction, are negligible \cite{novotny2006}. In such cases, the sample has to be rotated by \SI{90}{\degree}, see e.g.\,\cite{Sarau2014,Heilmann2016,Pezzotti2011}. Rotating the sample is, however, sometimes infeasible or even impossible. For example, in case of tiny crystal structures grown on a substrate, only structures that are close to the edge of the substrate can be accessed experimentally for titled illumination \cite{Sarau2014}. Certainly, the tilted substrate is disadvantageous even for these structures since it breaks the symmetry of the system and therefore may lead to an unwanted influence from the substrate. For these and many other reasons, it would be desirable to provide for more flexible excitation schemes, effectively overcoming the aforementioned limitation and offering a wide range of interaction scenarios with a single excitation beam impinging normally. Hence, no tilting of the sample or sophisticated sample preparation would be required anymore. This can be achieved by utilizing light beams exhibiting a significantly strong longitudinal electric field component, i.e., a field oscillating along the direction of propagation of the beam, at the position of the sample beside the conventional transverse components. By tailoring the electromagnetic field distribution at the nanoscale, a sub-wavelength-sized nanostructure my interact with different field configurations depending on its relative position with respect to the engineered field. A very practical and versatile route for the creation of a strong longitudinal component of the electric field is provided by tight focusing of, e.g., a \RPBlong \cite{Quabis2000,Youngworth2000,Dorn2003,Bauer2014}. Besides the strong longitudinal electric field component on the optical axis of a high numerical aperture objective lens, the focal field also features a ring-like distribution of transverse (in-plane) electric fields around the optical axis as it will be discussed in detail later on. Owing to their spatial degree of freedom, such light beams tailored at nanoscale dimensions have been proven already to be versatile tools in the linear or nonlinear study of individual nanostructures \cite{Kindler2007,Zuechner2008,Volpe2009,Banzer2010,Bautista2012,Bautista2013,wozniak2015}, in microscopy and imaging \cite{Biss2006,Hao2010}, Angstrom-localization of nanoparticles \cite{Neugebauer2016,Bag2018}, beam steering and guiding \cite{Neugebauer2014,Nechayev2019}, and the highly efficient coupling of individual photons to ions \cite{Golla2012}, to only name a few.\\
In the context of Raman spectroscopy, the utilization of tightly focused structured light beams is exceedingly rare. The few examples reported in the literature to date include, e.g., studies by Saito et al. \cite{Saito2008,Saito2012}, using tightly focused linearly polarized Gaussian beams -- which also feature a significantly strong longitudinal electric field component -- as well as radially and azimuthally polarized light beams focused with different numerical apertures onto homogeneous layers of Raman-active materials \cite{Saito2008,Saito2012}. The influence of the longitudinal electric field component on the recorded Raman spectra was studied. In another article, Roy et al. tightly focused radially polarized light to efficiently excite a tip in a TERS system \cite{Roy2010}. Recently, also paraxial spatial modes of light carrying orbital angular momentum were utilized for Raman spectroscopy \cite{Forbes2019}.

\section{Towards polarization-tailored Raman spectroscopy}
Inspired by the ideas discussed before, we propose and show with a proof-of-principle experiment a novel method, extending the toolbox of Raman-based spectroscopy techniques. Our work utilizes \textit{polarization-based excitation tailoring for extended Raman spectroscopy} (\textit{PETERS}). We apply our technique to study a sub-wavelength-sized gallium nitride (GaN) nanostructure similar to those investigated recently in Ref. \cite{Sarau2014}. With this novel method, we can access Raman modes sensitive to the $z$-component (longitudinal) as well as the $x,y$-components (transverse) of the electric field by placing the GaN nanostructure in different regions of the structured excitation realized by \TFlong \RPBlong impinging normally to the sample substrate (along the $z$-axis). This approach extends the capabilities of conventional Raman spectroscopy because no tilting of the sample or any complex sample preparation is required anymore to get access to the full variety of Raman modes in nanostructured Raman-active specimens. By choosing a sample of sub-wavelength dimensions, it is possible to record Raman modes excited by the $z$-component of the electric field when the GaN-sample is positioned on the optical axis. On the other hand, when the particle is displaced away from the optical axis, Raman spectra arising from the interaction with transverse focal field components ($x$- or $y$-components) can be observed.

\section{Experimental details and results}
\subsection{Experimental setup}
\begin{figure}[t]
    \centering
    \includegraphics[width=0.5\textwidth]{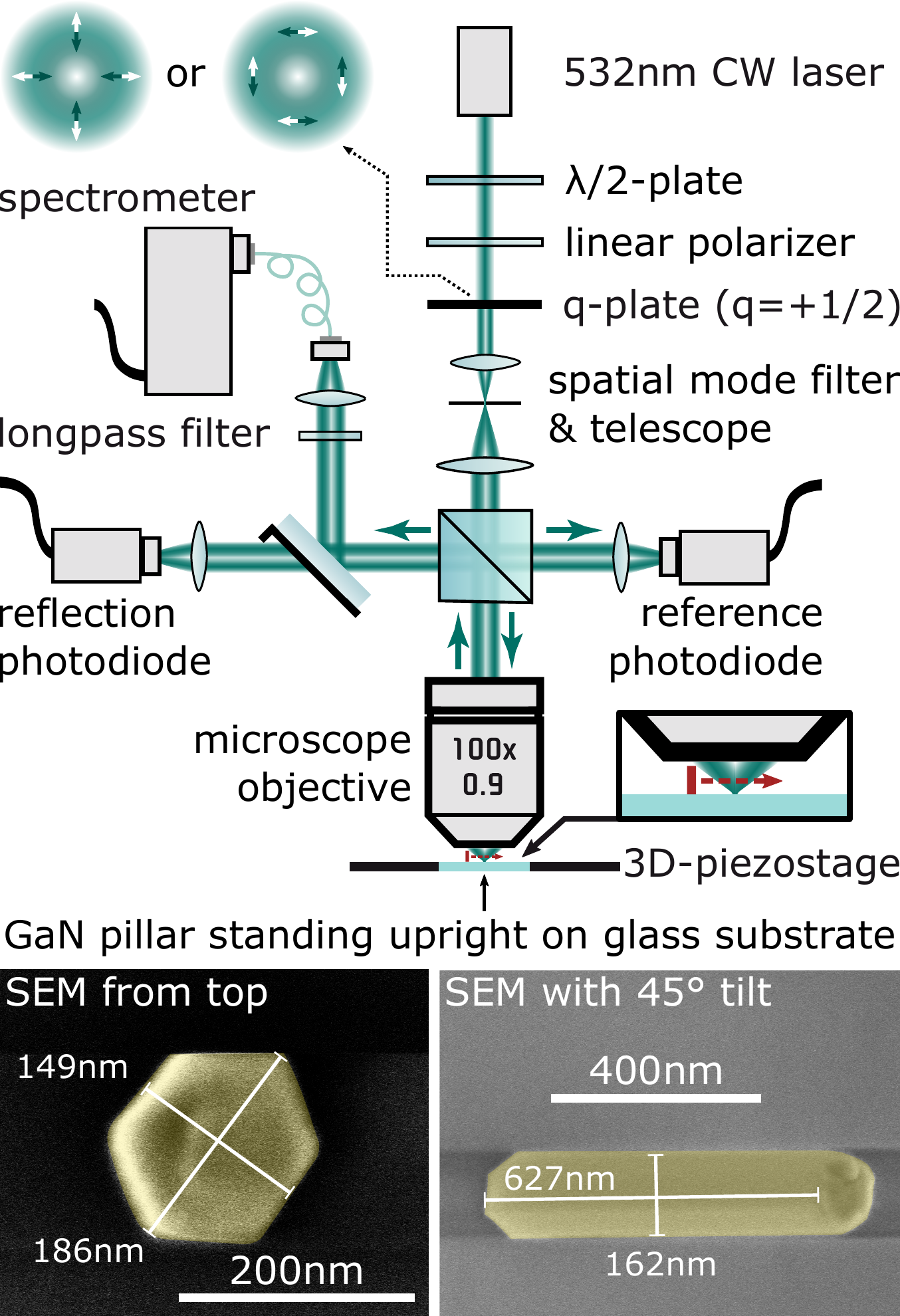}
    \caption{Top: Simplified sketch of the experimental setup utilized for our proof-of-principle experiment introducing polarization-based excitation tailoring for extended Raman spectroscopy. Beam generation, focusing, sample and detection stages are shown. Bottom: Scanning electron micrographs of one of the investigated GaN nano-pillars (colored in yellow for better visibility). Left: View along the crystal c-axis. Right: perspective view (45$^\circ$) indicating the height of the investigated structure. Optical measurements are performed for a non-tilted setting as mentioned in the text.}
    \label{fig:exp_setup}
\end{figure}

For the experimental proof of our proposed method, we use a custom-built setup similar to the ones discussed in more detail in \cite{Banzer2010, wozniak2015}. However, it should be noted here that the technique, in principle, could also be implemented in most standard commercial Raman system. A simplified sketch of our experimental system is depicted in Fig.\,\ref{fig:exp_setup} (top).
The light beam from a cw laser emitting at a wavelength of \SI{532.7}{nm} is converted to a cylindrical vector beam (radially or azimuthally polarized) using a combination of a half-wave plate, a linear polarizer and a polarization converter (q-plate) similar to the one introduced in \cite{marrucci2006}. The mode quality is increased with the help of a Fourier (spatial) filter consisting of two lenses and a pinhole. At the same time, the filter also acts as a telescope to adjust the beam diameter. A part of the beam is deflected onto a photo-diode using a non-polarizing beam-splitter. The measured signal is used for power normalization purposes. In the main beam path, the light is coupled into a microscope objective with high numerical aperture (NA = 0.9) and tightly focused onto the sample (see details below). The sample itself is mounted on a 3D-piezo-stage. Reflected light (substrate) as well as the light back-scattered elastically and inelastically by the individual Raman-active nanostructure is collected and collimated again by the same objective. A part of the collected light is guided by the aforementioned beam-splitter onto a flip-mirror. Subsequently, the beam is either directed onto a photo-detector for linear measurements or transmitted through a long-pass filter and coupled into a multimode fiber directing the light into a sensitive spectrometer.

\subsection{Sample fabrication and measurement procedures}
\begin{figure}[t]
    \centering
    \includegraphics[width=0.5\textwidth]{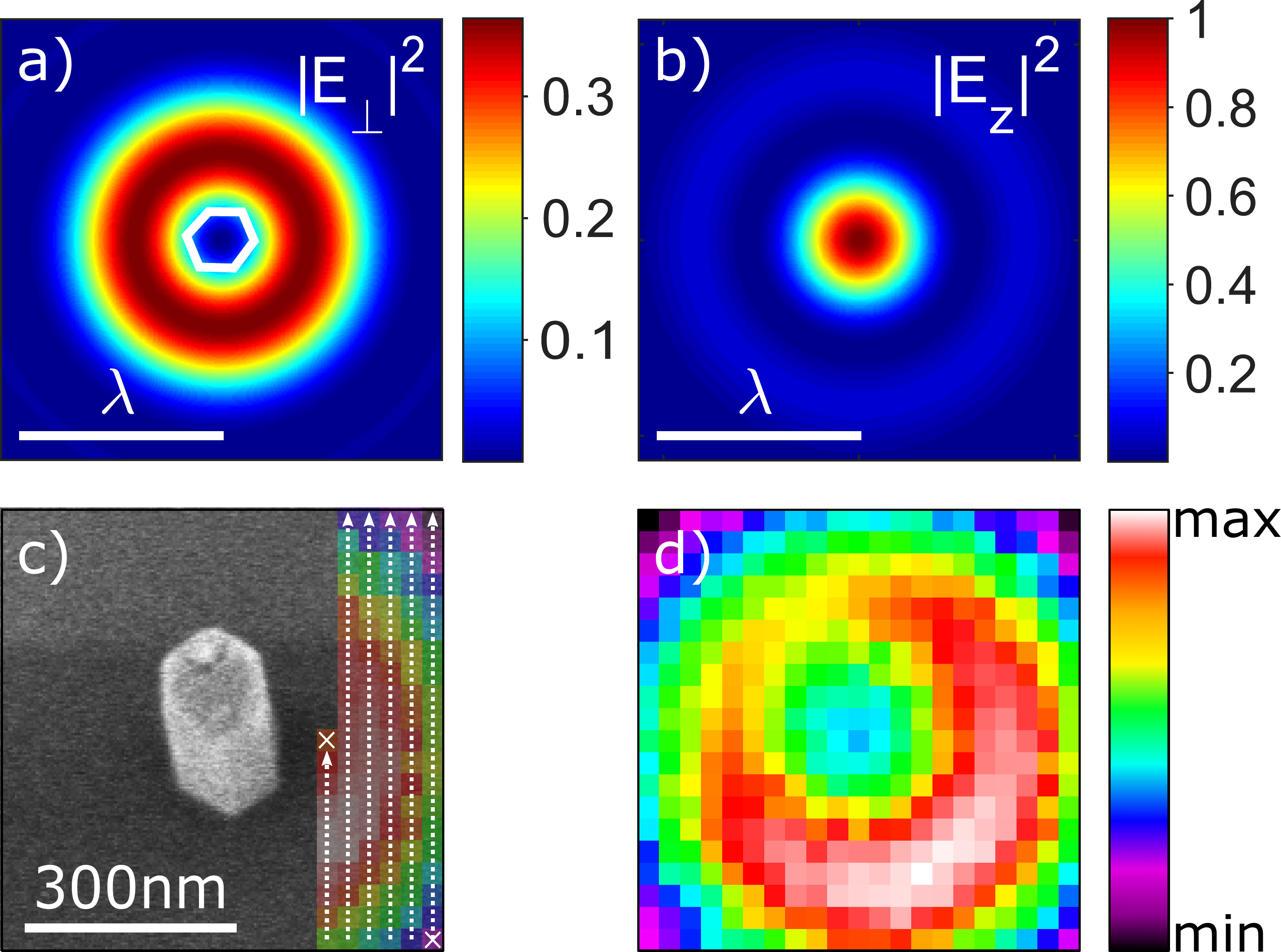}
    \caption{Focal-plane field distribution of a tightly focused radially polarized beam of wavelength $\lambda=532\,$nm, calculated using vectorial diffraction theory \cite{Richards1959}. a) Lateral field components $\left|\mathbf{E_\perp}\right|^2=\left|E_\mathrm{x}\right|^2+\left|E_\mathrm{y}\right|^2$ with the hexagon-like white contour representing the cross-section of the investigated GaN pillar (drawn to scale). b) Longitudinal field component. c) Top-view (SEM-image) of the GaN pillar with a partial reflection scan as overlay. In the experiment, the sample is scanned across the fixed beam using a piezostage. d) Full reflection scan corresponding to the region shown in c).}
    \label{fig:donut_focus}
\end{figure}

The sample studied here has been fabricated by metal-organic vapor phase epitaxy (MOVPE). The growth process on c-plane sapphire as well as the resulting GaN pillar-like nanostructures are equivalent to those discussed in detail in Refs. \cite{Tessarek2011,Tessarek2013,Tessarek2014}. For our proof-of-principle experimental study, we transferred the GaN pillars to a glass substrate (BK7) by a simple mechanical scraping approach. However, it should be noted here that this step is not necessary to make our proposed concept applicable. The transfer was only done for practical purposes. Using a scanning electron microscope (SEM), nano-pillars of sub-wavelength diameter which are standing upright on the substrate were identified. In Fig.\,\ref{fig:exp_setup}\, (bottom), we show two scanning electron micrographs of the investigated GaN nanostructure on the glass substrate. The height of the GaN nano-pillar is determined to be $\SI{627}{nm}/\sin(\ang{45}) = \SI{887}{nm}$ and its diameter was measured to be $\SI{149}{nm}$ to $\SI{186}{nm}$ (see hexagonal cross-section in Fig.\,\ref{fig:exp_setup}, bottom left). Its lateral dimensions are are therefore significantly smaller than the width of the focal field distribution used for excitation (as shown in Fig.\,\ref{fig:donut_focus}\,), enabling the realization of position-dependent interaction schemes.\\
Before performing Raman measurements, we first record linear reflection scans to localize the structure and find its position with respect to the focused field, i.e., the optical axis of the focusing objective. For completeness, we show the calculated focal field distribution of a tightly focused radially polarized light beam with a wavelength of 532\,nm, and all other parameters chosen equal to those used in the experiment, in Fig.\,\ref{fig:donut_focus}\,a)-b).\\

In the experiment, we scan the sample placed in the focal plane across the excitation beam using the aforementioned piezo-stage, while detecting the back-scattered and reflected light with a photo-detector. From the collected data, a raster-scan image is composed. Each pixel corresponds to the total power measured for a fixed position of the nanostructure in the beam and the given solid angle defined by the NA of the objective. The scan images show a distribution similar to the focal electric energy density distribution, owing to the sub-wavelength lateral dimensions of the GaN nanostructure. For a more in-depth discussion on the interpretation of such scan images, see \cite{Bauer2014,wozniak2015}. We schematically illustrate the measurement procedure and the relative sizes of the beam in the focal plane and the studied nanostructure in Fig.\,\ref{fig:donut_focus}\,a)-c).\\
In Fig.\,\ref{fig:donut_focus}\,d), we show a scan image recorded in reflection for such a radially polarized input beam. In the ideal of a cylindrically symmetric beam and nanostructure, also the scan image should be cylindrically symmetric. The observed asymmetries of the measured scan image can be attributed to the elongated cross-section of the investigated nano-pillar, its slightly tilted configuration and minor beam imperfections. The corners of the scan image correspond to particle positions outside the beam where the beam does not overlap with the structure anymore. For such positions, only the reflection from the substrate is measured. For the nanostructure placed on-axis (center of the depicted scan image), a signal different from the background reflection is recorded due to the strong on-axis longitudinal electric field component. The stronger signal observed for off-axis in comparison to on-axis particle positions has multiple causes, including the excited modes in the particle as well as the collection geometry of the system as discusse din more detail below. This linear scan measurement already indicates that a position-dependent interaction of the focal fields with the nanostructure can be realized with the chosen excitation beam (similar to previously reported linear investigations \cite{Banzer2010,Bauer2015,wozniak2015}). However, the aforementioned scan only serves for alignment purposes in this study, and for illustrating the spatial dependence of the light-matter interaction.

\subsection{Theoretical description of Raman scattering}
\subsubsection{Conventional illumination and collection geometries}
\label{sssec:raman_tensors}
In conventional Raman spectroscopy, the impinging light field is typically linearly polarized with a polarization direction defined by the electric field $\mathbf{e}_\mathrm{inc}$. Usually, a second polarizer transmitting an electric field parallel to $\mathbf{e}_\mathrm{sca}$ is introduced as an analyzer in the detection path. The Raman intensity of the measured scattered light is given by (see e.g. Refs.\,\cite[Eq.\,7]{Turrell1996} and \cite[Eq.\,36]{Loudon1964})
\begin{align}\label{eq:raman_int_alpha}
I_\mathrm{sca} \propto |\mathbf{e}_\mathrm{inc} \cdot \boldsymbol{\alpha} \cdot \mathbf{e}_\mathrm{sca}|^2,
\end{align}
where $\boldsymbol{\alpha}$ is the Raman tensor of order two. Gallium nitride (GaN) features a wurzite crystal structure belonging to the dihexagonal pyramidal point-group $C_{6v}$ (Schoenflies notation). It is a hexagonal crystal system and consists of two interpenetrating hexagonal closely packed sublattices, each with one type of atom, which are offset along the c-axis by $3/8$ of the unit cell height \cite{Morkoc2009}. The Raman tensors $\boldsymbol{\alpha}$ of the system inherent \ATO, \ETO and \Eh mode have the following form \cite{Kroumova2003}
\begin{align}\label{eq:Raman_tensor_GaN}
    \boldsymbol{\alpha}_\mathrm{\ETO} {=}
    \begin{psmallmatrix*}[r]
    0 & 0 & c\\
    0 & 0 & c\\
    c & c & 0
    \end{psmallmatrix*}
    ,\quad
    \boldsymbol{\alpha}_\mathrm{\ATO} {=}   
    \begin{psmallmatrix*}[r]
    a & 0 & 0\\
    0 & a & 0\\
    0 & 0 & b
    \end{psmallmatrix*}
    , \quad
    \boldsymbol{\alpha}_\mathrm{\Eh} {=}
    \begin{psmallmatrix*}[r]
    d & -d & 0\\
    -d & -d & 0\\
    0 & 0 & 0
    \end{psmallmatrix*},
\end{align}
using Cartesian coordinates such that the z-axis is parallel to the c-axis ([0001]-axis) of the wurzite GaN crystal (see \cite{Morkoc2009}). In the literature, the values for $a$, $b$, $c$ and $d$, which describe how strongly different polarization directions of the incoming light couple to different vibrational modes (see Eq.\,\ref{eq:raman_int_alpha}), are mostly determined empirically using experiments. However, also theoretical approaches have also been presented especially for small molecules \cite{Ligeois2005,Irmer2014,Pezzotti2011,Livneh2006}. The values of the Raman tensor values reported in the literature vary strongly. Comparing for example Refs.\,\cite{Pezzotti2011} and \cite{Irmer2014}, the ratio $a/b$ is different by a factor of $8$ and the value for the ratio $c/d$ differs by \SI{40}{\%}. Another publication, Ref.\,\cite{Livneh2006}, retrieves a ratio $a/b$ which differs by \SI{-15}{\%} to those in \cite{Irmer2014}.

\subsubsection{Structured illumination and high-NA collection}
Compared to conventional Raman systems, where Eq.\,\ref{eq:raman_int_alpha} is readily applicable because excitation and observation are done along a single direction only, we need to adapt this approach due to the structured and high NA illumination and collection we choose. In order to describe the intensity reaching the spectrometer, we need to calculate the overlap of the Raman scattered field (we assume dipolar emission) collected with the high-NA optics and the light transmitted through an analyzer (not shown in \ref{fig:exp_setup}) positioned in the collimated beam in the far-field. By integrating all analyzer positions, as detailed in \cite{Turrell1984,Saito2008}, we arrive at the Raman scattering intensities for the different Raman modes measured at the spectrometer without an analyzer in place.
\begin{subequations}\label{eq:ramanintresult}
    \begin{align}
    I_\mathrm{\ETO} &= \,\, c^2 B I_\perp + 2 c^2  A I_z, \label{eq:raman_int_result_E1TO}\\
    I_\mathrm{\ATO} &= \,\, a^2 A I_\perp + \,\, b^2 B I_z, \label{eq:raman_int_result_A1TO}\\
    I_\mathrm{\Eh} &= 2 d^2 A I_\perp, \label{eq:raman_int_result_E2h}
    \end{align}
\end{subequations}
with
\begin{align}\label{eq:integratedlongitudinalenergydensity}
I_\perp = I_x + I_y
 \,\,\mathrm{and }\,\, I_{\{ x,y,z\}} = \int_V\mathrm{d}V |E_{\mathrm{pillar}, \{ x,y,z\}}|^2
\end{align}
the integrated electric energy density components within the GaN pillar of volume $V$. The factors A and B arise from the geometry of the collection optics with numerical aperture $\mathit{NA}=\sin\theta_\mathrm{max}$ and are given by (see \cite{Turrell1984}; corrected in \cite[Eq.\,32]{Turrell1989})
\begin{subequations}\label{eq:AB_collection_formula}
    \begin{alignat}{2}
    A &=\phantom{2}\pi^2 \int_{0}^{\theta_\mathrm{max}}d\theta (\cos^2\theta+1)\sin\theta \nonumber\\
    &= \frac{\pi^2}{3}  \left( -\cos^3\theta_\mathrm{max} - 3 \cos\theta_\mathrm{max} +4 \right),\\ 
    B &=2\pi^2 \int_{0}^{\theta_\mathrm{max}}d\theta \sin^3\theta \nonumber\\
    &= \frac{2\pi^2}{3} \left( \cos^3\theta_\mathrm{max} - 3 \cos\theta_\mathrm{max} +2 \right). 
    \end{alignat}
\end{subequations}
Therefore, the ratio $A/B$ can be interpreted as a measure of the collection efficiency of the microscope objective for an electric dipole oscillating in free space perpendicular to the optical axis ($x$- or $y$-dipole) compared to a dipole oscillating parallel to the optical axis ($z$-dipole). For $NA\rightarrow 0$ ($\theta_\mathrm{max} \rightarrow 0$), $A$ and $B$ simplify to the case of conventional low-$\mathit{NA}$ Raman spectroscopy and $A/B\rightarrow\infty$. For $\mathit{NA} =1$ ($\theta_\mathrm{max}=\SI{90}{\degree}$), the full upper half-space is collected, resulting in $A=B$. Using a numerical aperture of our setup, $\mathrm{NA} = 0.9$ ($\theta_\mathrm{max} \approx \ang{64.16}$), results in the collection ratio $A/B \approx 1.68$.  Therefore, by using an microscope objective with an $\mathrm{NA}$ of $0.9$, the emission of a dipole oscillating perpendicular to the optical axis is observed stronger than a dipole oscillating parallel to the optical axis, partially also explaining the linear scan map shown in Fig.\,\ref{fig:donut_focus} d).

\subsection{Polarization-tailored Raman measurements}
\begin{figure*}[t]
    \centering
    \includegraphics[width=0.95\textwidth]{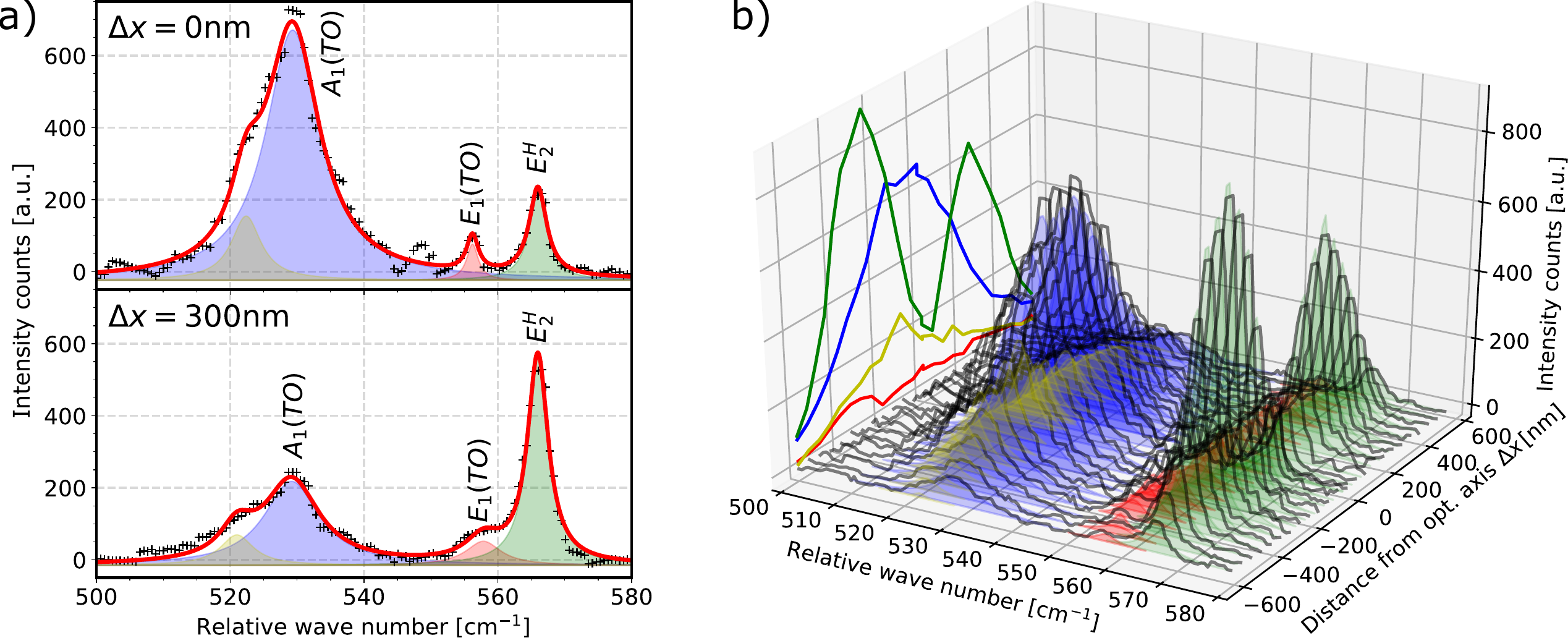}
    \caption{a) Raman spectra recorded using the radially polarized excitation scheme with displacements of $\Delta x=\SI{0}{nm}$ (top) and $\Delta x= \SI{300}{nm}$ (bottom) with respect to the optical axis. Black crosses: experimental data points; red line: baseline corrected fitted data achieved using an asymmetric least-square algorithm (see Ref. \cite{ALSbaseline}); colored areas: fitted Gaussian functions for retrieval of the Raman peak positions. b) Raman spectra recorded for displacements in \SI{50}{nm} steps along the $x$-axis. The projection plane to the left shows the position dependent amplitudes of the Gaussian functions fitted to the individual Raman peaks shown in the same colors as indicated in a).}
    \label{fig:radial_spectra}
\end{figure*}

Following the procedure discussed above, Raman spectra were recorded for different positions of the nano-pillar in the structured focal field. The data was baseline corrected using the asymmetric least-square algorithm described in Ref. \cite{ALSbaseline}. The baseline corrected spectral data was fitted by four Gaussian functions, to extract the relative intensities and wavenumbers of the Raman modes observed in the measurements. The resulting plots are shown in Fig.\,\ref{fig:radial_spectra}\,a) for the case of the GaN nano-pillar sitting on the optical axis (top) and the case of an off-axis position (\SI{300}{nm}). For both cases, the obtained values  of the three Raman modes described before are $529.09\pm0.26$\invcm (\ATO), $557.1\pm0.8$\invcm (\ETO) and $565.86\pm0.13$\invcm (\Eh). For the first case (on-axis), the height of the \ATO-peak is significantly higher than the \Eh-peak. For the second case (off-axis), the \ATO-peak is much weaker and the \Eh-peak is dominant. In contrast, the \ETO-peak is stronger if the GaN nano-pillar is on-axis. Two more peaks were observed in the recorded spectra. One is located at $521.5\pm0.5$\invcm. This additional feature near the \ATO mode is found in the recorded spectra for different particle positions and it contributes strongly. It is close to the characteristic Raman shift associated with bulk silicon (Si), although no silicon is expected to be present in the samples. Interestingly, the peak scales with the \ATO mode and is therefore expected to be attributed to a surface phonon carrying less energy  than the\ATO oscillations in the bulk medium. Further experiments are necessary to validate this hypothesis. Furthermore we observed additional peaks occurring in the spectral region between the \ETO and the \ATO peaks, see for instance the peak at $548$\invcm in Figure \ref{fig:radial_spectra}\,a). The spectral position of these peaks varied for different positions within the exciting focal field. These modes could be attributed to so called quasi-TO modes similar to those observed in Ref. \cite{Sarau2014}. Their examination may reveal additional interesting features upon further investigation. However, their detailed investigation goes beyond this proof-of-concept study and since they do not influence the analysis of the main peaks, they were neglected in this study.

\begin{figure*}[t]
    \centering
    \includegraphics[width=\textwidth]{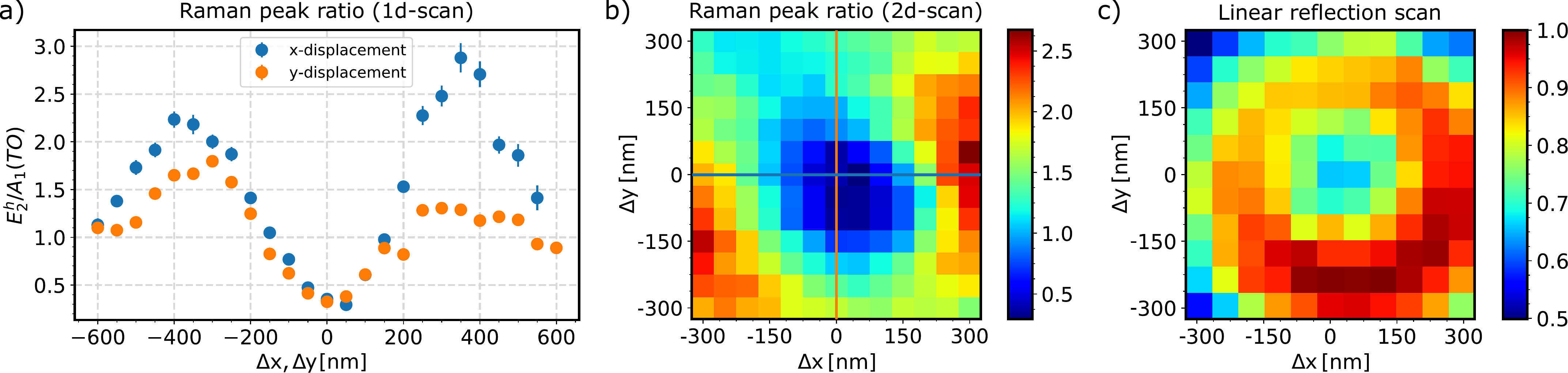}
    \caption{a) Ratio of the Raman peak intensities of \Eh/\ATO shown for displacements in $x$- and $y$-direction. b) Same ratio as plotted in a) but now for a 2D scan. The orange and blue lines corresponging to the line-scans in a). c) Normalized reflection scan (corresponds to Figure \ref{fig:donut_focus} d)).}
    \label{fig:radial_ratios}
\end{figure*}

The spectra were also measured for several other displacements from \SI{-600}{nm} to \SI{600}{nm} with the corresponding data being summarized in Figure\,\ref{fig:radial_spectra}\,b). We find that the spectra smoothly change with the displacement. The height of the \ATO peak steadily decreases when moving off the optical axis whereas the height of the \Eh peak increases until it reaches its maximum at a displacement of roughly \SI{\pm300}{nm}. Beyond this lateral position, also the \Eh peak decreases. In the following discussion and we will concentrate on the two main Raman modes, i.e.,  the \ATO and the \Eh modes.\\
The aforementioned observations can be explained based on the focal structure of the field interacting with the GaN nano-pillar. When the GaN nano-pillar is moved off the optical axis, the \Eh peak increases because the transverse electric field overlaps more and more strongly. It is maximum for an $x$- or $y$-displacement of approximately \SI{300}{nm}, in good agreement with the positions of the maximum transverse and zero longitudinal electric field components in the focus as indicated in Fig.\,\ref{fig:donut_focus}\,a). Similarly, the \ATO peak is strongest when the GaN nano-pillar is placed on-axis since the longitudinal and total electric energy density are strongest on the optical axis. For a more quantitative investigation, pre-factors $a$, $b$ and $c$ as well as $A$ and $B$ in Eq.\,\ref{eq:ramanintresult} need to be taken into account. Since in the literature the values for $a$, $b$ and $c$ vary strongly as discussed above, we continue our analysis on a general level without basing it on specific values. With the help of Equations \ref{eq:raman_int_result_A1TO} and \ref{eq:raman_int_result_E2h}, we find that the ratio of \Eh and \ATO peaks is directly linked to the ratio of the energy densities of the longitudinal and the transverse components of the electric field,
\begin{align}
\label{eq:I_E2h_frac_I_ATO}
    \frac{I_\mathrm{\Eh}}{I_\mathrm{\ATO}} &= \frac{2 d^2}{a^2+b^2 \frac{B}{A} \frac{I_z}{I_\perp}}.
\end{align}
The measured ratios are shown in Figure\,\ref{fig:radial_ratios}\,a) and b). We can see that it is minimum on the optical axis and increases up to its maximum at a displacement of approximately \SI{300}{nm}. For larger displacements, it decreases again. Besides minor asymmetries to be discussed later on, the curves are comparable for positive and negative $x$- and $y$-displacements as one expects for symmetry reasons. As we depicted in Fig.\,\ref{fig:donut_focus}, the longitudinal electric field is significantly stronger on-axis. Therefore, the \ATO mode is excited stronger with respect to the \Eh mode and therefore the ratio in Eq.\,\ref{eq:I_E2h_frac_I_ATO} is lowest. As soon as the GaN nano-pillar is moved off-axis, the height of the \Eh peak increases relative to the \ATO peak since the transverse electric field overlaps more and more with the GaN nano-pillar and the longitudinal electric field is decreasing quickly. A larger contribution of the transverse field increases the ratio in  Eq.\,\ref{eq:I_E2h_frac_I_ATO}. The observed asymmetries are similar to those for the linear reflection measurement depicted again for comparison in Fig.\,\ref{fig:radial_ratios}\,c). This could be attributed to imperfections in the distribution of the excitation beam. This could be overcome partially by using polarization-preserving beam-splitter-pairs not influencing the polarization and intensity distribution of the input beam (see Refs. \cite{Tidwell1992,Banzer2010,wozniak2015}). In addition, the GaN nano-pillar is not cylindrically symmetric due to its crystal structure and it is slightly tilted. This kind of asymmetry can cause a different response for particle displacements along corresponding axes, see Figsure\,\ref{fig:exp_setup} and \ref{fig:donut_focus}, also highlighting the potential capabilities of \textit{PETERS}.

\subsection{Towards quantitative measurements and data analysis}
In Fig.\,\ref{fig:radial_spectra} a), it can be seen that the \Eh mode is rather strongly excited for an on-axis placement of the nano-pillar, even though the geometric overlap of the GaN nano-pillar with the transverse electric field is actually very weak (compare Fig. \ref{fig:donut_focus} a)). To explain this peculiarity, we need to consider not only the exciting electric field geometrically overlapping with the GaN nano-pillar but the actual electric field components inside the GaN nano-pillar resulting from the light-matter interaction as briefly mentioned above (see Eq. \ref{eq:integratedlongitudinalenergydensity}). Their distribution can be quite complex and will inherently feature both transverse and longitudinal electric field components owing to the nature of the photonic modes excited in the nano-pillar. To reveal the actual mode structure, we performed 3D finite difference time domain (3D-FDTD) simulations.\\

\begin{figure}[ht]
    \centering
    \includegraphics[width=0.75\textwidth]{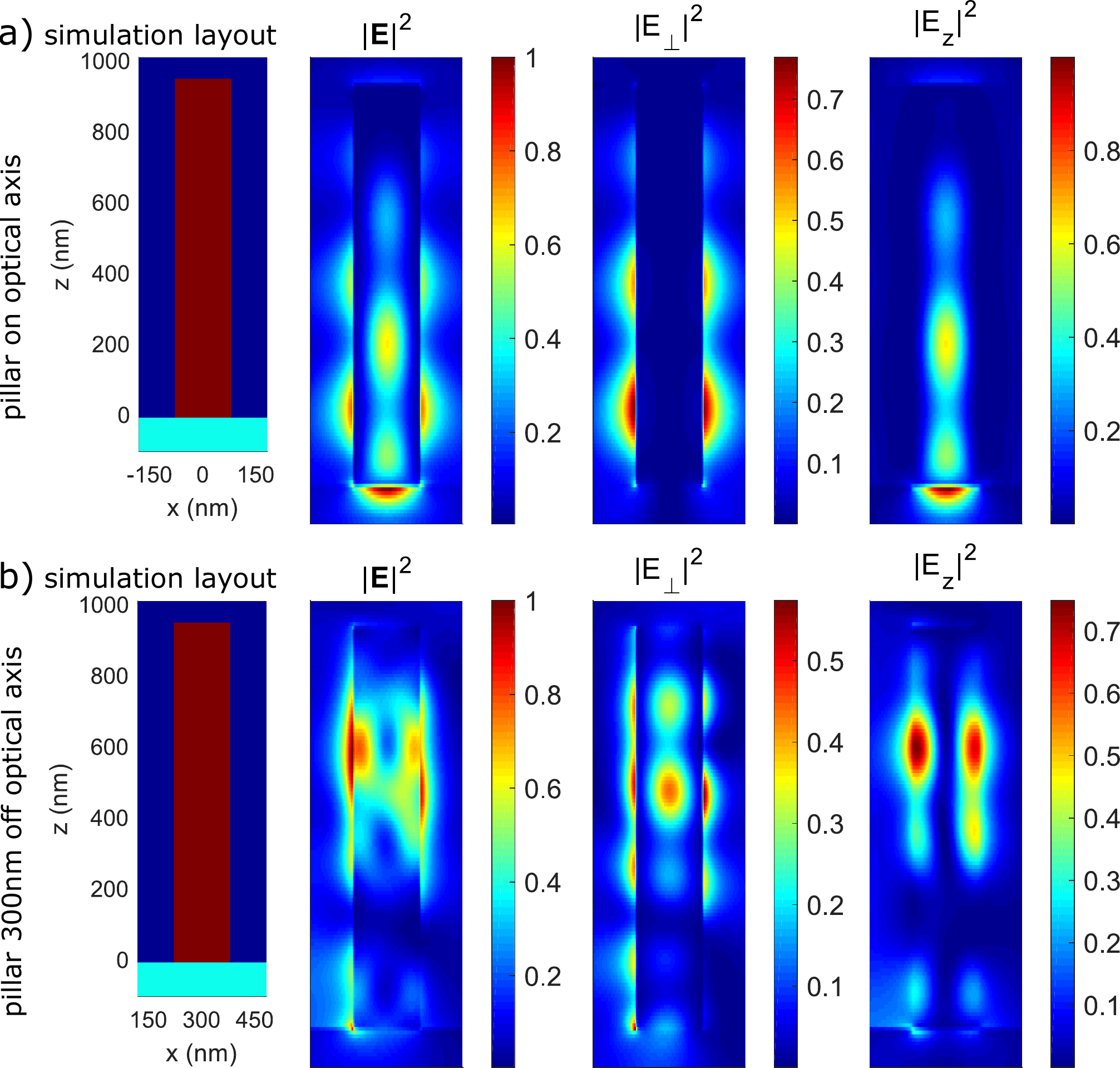}
    \caption{Near-fields extracted from 3D-FDTD simulations. Left: Cross-sectional view of the simulation layout with the excitation beam impinging from top and the nano-pillar (red) placed on substrate (light blue). Right: Total as well as transverse and longitudinal electric energy densities in the defined cross-section.  a) Simulation with the pillar placed centrally on the optical axis. b) Simulation with the pillar displaced by \SI{300}{nm} away from the optical axis.}
    \label{fig:FDTD_nearfields}
\end{figure}

In Figure \ref{fig:FDTD_nearfields}, the numerically calculated near-fields are shown in an $xz$-cross-section cutting centrally through the pillar. The excitation beam was calculated using vectorial diffraction theory and implemented as a source in the numerical simulations. Similar to the experiment, the focal plane is placed \SI{300}{nm} above the substrate ($z$ = $300$nm). In Figure \ref{fig:FDTD_nearfields}\,a), the near-fields for the nano-pillar placed on-axis are shown. In Figure \ref{fig:FDTD_nearfields}\,b), the corresponding data is presented for a lateral displacement of \SI{300}{nm} along the positive $x$-axis, where the lateral field components reach their maximum value. As one can see, the arising mode structure is complex. However, the main field contributions in the simulations underline the objective of our idea, i.e., position-dependent coupling and Raman spectra. When the GaN pillar is positioned on the optical axis, the longitudinal electric field component inside the pillar is dominant and the transverse components are rather weak. Nonetheless, they are non-zero and contribute to the \Eh line. However, with the GaN pillar positioned off-axis, the longitudinal electric field components are not negligible due to the formation of a complex mode structure. Although the sub-wavelength structure overlaps dominantly with a transverse electric field for this off-axis position, the actual Raman signal is ruled by the electric fields forming inside the pillar. Hence, the Raman spectra observed for this particle placement feature contributions from both transverse and longitudinal field components. It should also be noted here that the far-field scattering cannot be described appropriately by a simple dipole approximation because the modes are of higher order. The excited modes influence the aforementioned parameters $A$ and $B$ and also depend on the excitation wavelength, hence also affecting the collection efficiencies. These aspects need to be taken into account in future steps, enabling also a precise quantitative analysis.\\
Our experimental and numerical data show unambiguously that we can indeed selectively excite the Raman-active nanostructure under investigation by taking advantage of the tailored focal fields, here demonstrated with a tightly focused radially polarized cylindrical mode. The input field can be tailored to feature more sophisticated focal field distributions enabling many more interaction schemes. As indicated above, the comparison of Raman spectra for different particle positions can pave the way towards a quantitative analysis of the recorded data. Furthermore, the collected reflection signal can be filtered in momentum space to emulate the detection within different solid angles (similar to \cite{Bauer2014}). This way, the free parameters in the Eq. \ref{eq:raman_int_result_E1TO}-\ref{eq:raman_int_result_E2h} above could be retrieved quantitatively. To interpret correctly the position-dependent signals, also the actual modes excited in the nanostructure need to be taken into account as indicated by the numerical results in Fig. \ref{fig:FDTD_nearfields}. An experimental far-field multipolar analysis (see, e.g., \cite{Eismann2018}) can help in identifying the modes, also making possible a quantitative analysis of the Raman spectra recorded with \textit{PETERS}.

\section{Conclusion}
In conclusion, we have proposed, and experimentally investigated in a proof-of-concept study, a novel potential addition to the toolbox of Raman spectroscopy. The underlying \textit{polarization-based excitation tailoring for extended Raman spectroscopy} (\textit{PETERS}) takes advantage of the spatial structure of tightly focused vector beams for recording position-dependent Raman spectra within a single beam scenario. Hence, the three-dimensionally structured light field grants access to multiple excitation schemes of Raman modes without the necessity for tilting the sample or requiring any exotic sample preparation. To the best of our knowledge, this is the first experimental demonstration of polarization-tailored Raman spectroscopy of Raman-active nanostructures, capitalizing the spatial degree of freedom of structured light. In combination with a more quantitative and detailed analysis discussed briefly above, our study might pave the way for the implementation of \textit{PETERS} in commercial Raman systems. In addition, this method might also be applicable in TERS, where the excited Raman modes in a nano-sized structure under study also depend on its relative position to the tip used for field enhancement, owing to the inherently structured nature of near-fields. By this step, the enhancement of scattering efficiencies provided by TERS together with the accessibility of multiple excitation schemes and Raman modes in a single excitation field scenario facilitated by \textit{PETERS} could be combined.

\section{Funding}
We acknowledge financial support by the Max Planck - University of Ottawa Centre for Extreme and Quantum Photonics in Ottawa, Canada. GS and SC acknowledges financial support by the German Research Foundation (DFG) within the research projects Dynamics and Interactions of Semiconductor Nanowires for Optoelectronics (FOR 1616) and Hybrid Inorganic/Organic Systems for Opto-Electronics (HIOS, SFB 951).

\section{Acknowledgements}
We thank Christian Tessarek and Martin Heilmann for fabricating the raw GaN nanostructures on sapphire. We thank Johannes Zirkelbach and Stephan G\"otzinger for their help with the spectrometer.

\bibliography{literaturlistgrosche}

\end{document}